\documentclass[square,sort,comma,numbers]{article}
\usepackage[utf8]{inputenc}

\title{Color induction in equiluminant flashed stimuli}
\author{Xim Cerda-Company and Xavier Otazu\\ Computer Vision Center, Computer Science Department,\\ Universitat Autonoma de Barcelona, Spain}
\date{June 2018}

\usepackage{graphicx}
\usepackage{amssymb}
\usepackage{amsthm}
\usepackage{amsmath}
\usepackage{amsfonts}
\usepackage{lineno}
\usepackage{textcomp}
\usepackage{natbib}
\usepackage{arydshln}
\usepackage{multirow}
\usepackage{multicol}
\usepackage{subcaption}
\usepackage{gensymb}
\usepackage{arydshln}

\usepackage{natbib}
\usepackage{graphicx}

\begin{document}

\maketitle

\begin{abstract}
Color induction is the influence of the surrounding color (inducer) on the perceived color of a central region. There are two different types of color induction: color contrast (the color of the central region shifts away from that of the inducer) and color assimilation (the color shifts towards the color of the inducer). Several studies on these effects used uniform and striped surrounds, reporting color contrast and color assimilation, respectively. Other authors (Kaneko and Murakami,~\textit{J Vision}, 2012) studied color induction using flashed uniform surrounds, reporting that the contrast was higher for shorter flash duration. Extending their work, we present new psychophysical results using both flashed and static (\textit{i.e.}, non-flashed) equiluminant stimuli for both striped and uniform surround. Similarly to them, for uniform surround stimuli we observed color contrast, but we did not obtain the maximum contrast for the shortest ($10\, ms$) flashed stimuli, but for $40\, ms$. We only observed this maximum contrast for red, green and lime inducers, while for a purple inducer we obtained an asymptotic profile along flash duration. For striped stimuli, we observed color assimilation only for the static (infinite flash duration) red-green surround inducers (red 1st inducer, green 2nd inducer). For the other inducers' configurations, we observed color contrast or no induction. Since other works showed that non-equiluminant striped static stimuli induce color assimilation, our results also suggest that luminance differences could be a key factor to induce it.
\end{abstract}

\section{Introduction}
The color appearance of objects can be influenced by the color of the surrounding objects. This effect is called color induction and it can be classified in two different types: color contrast and color assimilation. Color contrast occurs when the perceived color of the object shifts away from the surrounding color (inducer), while color assimilation occurs when the perceived color shifts toward the inducer. Both color contrast and color assimilation have been studied since the 19th century~\cite{Chevreul1839,VonBezold1876}. Although assimilation is more common than contrast in daily life~\cite{DeValois1988}, contrast has been more widely studied, mainly using static stimuli (\textit{i.e.}, the stimuli did not change along time).

Several authors~\cite{Monnier2003,Monnier2004,Otazu2010} have shown that, in general, uniform surrounds induce color contrast and striped surrounds tend to induce color assimilation. In striped surround stimuli, spatial frequency of stripes is a key factor to induce color assimilation~\cite{Fach1986,Smith2001}, with higher spatial frequencies ($>2.74\, cpd$) leading to a stronger assimilation~\cite{Cao2005,Otazu2010}. Nevertheless, Smith et al.~\cite{Smith2001} found that thick ($<0.7\, cpd$) stripes can induce color contrast.

Color induction has also been studied using dynamic and flashed stimuli~\cite{Anstis1978,Kelly1993,DeValois1986,Singer1994,Kaneko2012}. In dynamic stimuli, the inducer is modulated along time, being the temporal frequency of the surround modulation an important factor for color induction, \textit{e.g.} stronger induction at low temporal frequencies, falling down beyond $2-3\, Hz$~\cite{DeValois1986,Singer1994}. In flashed stimuli, the target stimulus is presented during a brief time (a 'blank' frame is usually shown when the target stimulus is not presented)~\cite{Kaneko2012}. Some of these studies measured the color induction of afterimages~\cite{Anstis1978,Kelly1993}. They showed that color contrast can produce afterimages and, conversely, color afterimages can induce color contrast~\cite{Anstis1978}. Furthermore, Kelly and Martinez-Uriegas~\cite{Kelly1993} concluded that isoluminant chromatic stimuli create isoluminant chromatic afterimages.

Recently, Kaneko and Murakami~\cite{Kaneko2012} published an extensive work in color induction using equiluminant flashed color stimuli with uniform surrounds. They measured the color induction at different flash durations (from $10\, ms$ to $640\, ms$) and showed that color contrast significantly depends on the duration of the flash. They concluded that the shorter the flash duration, the stronger the contrast. Since they used uniform surrounds, only color contrast was reported.

\begin{figure}[htb!]
    \begin{subfigure}[b]{.48\linewidth}
      \centering
      \centerline{\includegraphics[width=0.7\linewidth]{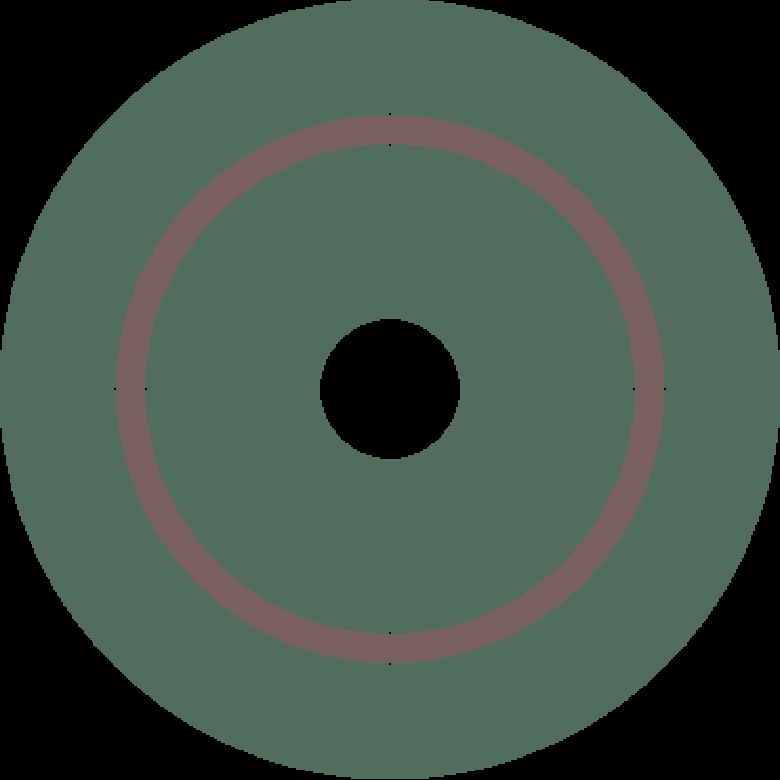}}
	\captionsetup{justification=centering}
    \end{subfigure}
    \hfill
    \begin{subfigure}[b]{0.48\linewidth}
      \centering
      \centerline{\includegraphics[width=0.7\linewidth]{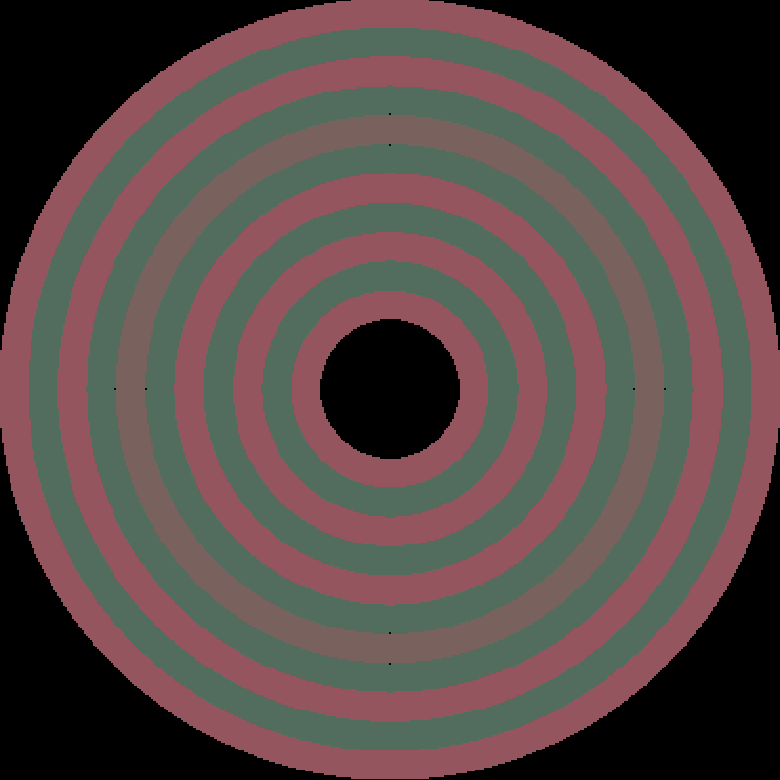}}
	\captionsetup{justification=centering}
    \end{subfigure}
    \begin{subfigure}[b]{.48\linewidth}
      \centering
      \centerline{\includegraphics[width=0.7\linewidth]{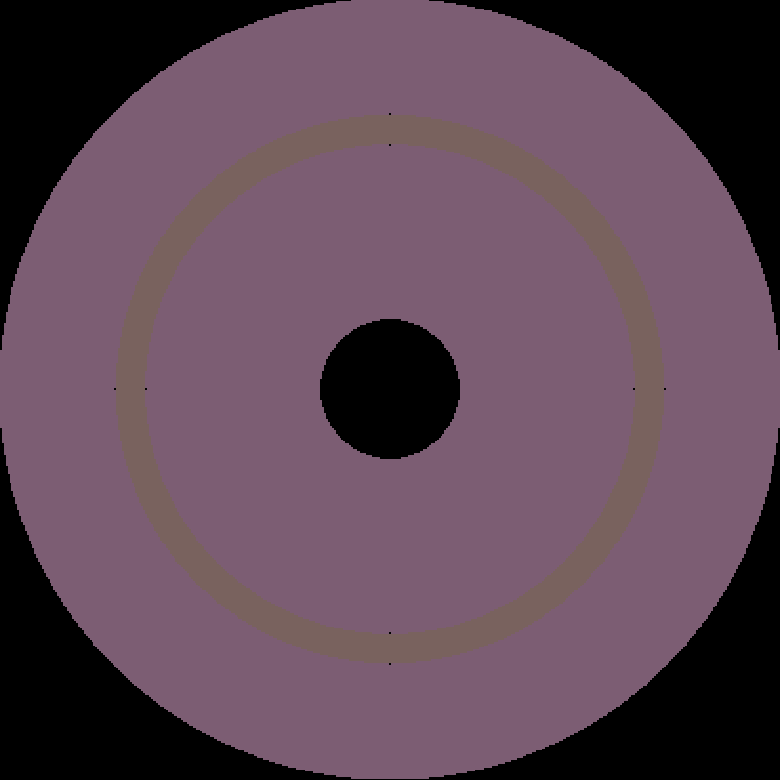}}
	\captionsetup{justification=centering}
    \end{subfigure}
    \hfill
    \begin{subfigure}[b]{.48\linewidth}
      \centering
      \centerline{\includegraphics[width=0.7\linewidth]{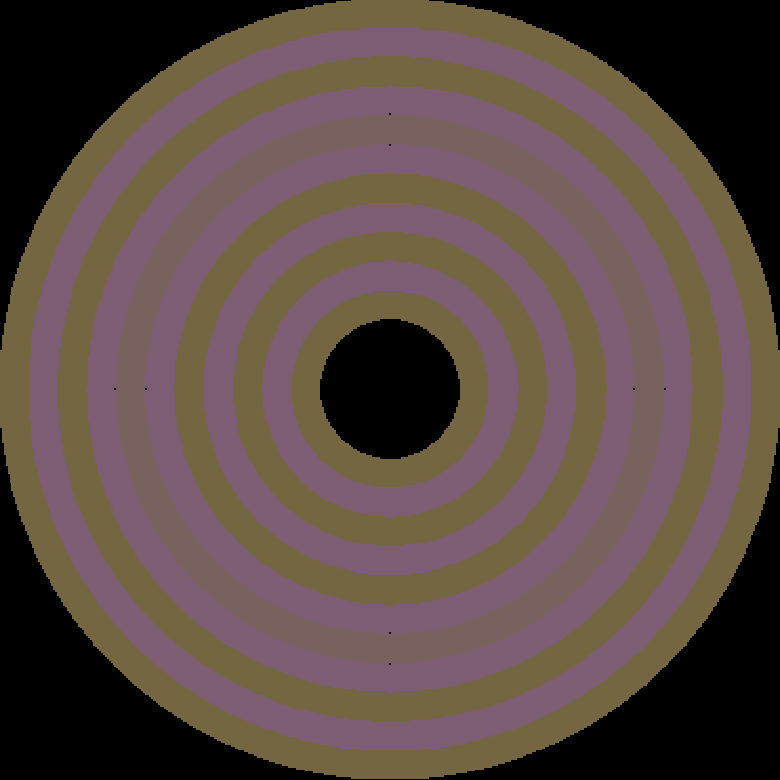}}
	\captionsetup{justification=centering}
    \end{subfigure}
    \caption{Examples of some stimuli used in this work (see Table~\ref{table:conditions}). In this figure, we can observe that the achromatic ring (surrounded by 8 black dots) is perceived differently depending on the surround chromaticity. This effect is called color induction. For instance, on the top-left figure, the achromatic ring surrounded by a green inducer is perceived as reddish, while the achromatic ring surrounded by a purple inducer (bottom-left figure) is perceived as lime-ish.}
    \label{fig:spatialConfigExamples}
\end{figure}

In this work, we extended Kaneko and Murakami's work~\cite{Kaneko2012}, presenting a new study of color induction using both uniform and striped surrounds and both static and flashed stimuli (see several static stimuli examples in Figure~\ref{fig:spatialConfigExamples}). Similarly to other works ~\cite{Monnier2003,Monnier2004,Otazu2010,Cao2005,Fach1986,Smith2001,Kaneko2012,Anstis1978,Kelly1993,DeValois1986,Singer1994}, we expect to observe color contrast in uniform surround stimuli and color assimilation in striped surround stimuli. Furthermore, we expect to reproduce the Kaneko and Murakami's results~\cite{Kaneko2012} for flashed uniform surrounds and to analyze whether color induction depends on flash duration for both striped and uniform surrounds. In previous papers~\cite{Otazu2008,Otazu2010,Penacchio2013} one author of this paper simultaneously reproduced psychophysical results of both color and brightness induction using a Wavelet model and a neurodynamical model of V1. These models suggest that color contrast and color assimilation could be the result of the same mechanism (lateral connections)~\cite{Zaidi1992,Zaidi1999,Penacchio2013}.

Thus, although uniform and striped surrounds could induce opposite effects (color contrast and color assimilation, respectively), our hypothesis is that, when the chromatic surrounds are flashed, the temporal behavior of the induced color would be similar in both cases.

\section{Methods}
\label{sec:methods}

\subsection{Apparatus}
\label{sec:apparatus}
All experiments were conducted in a dark room, on a calibrated $21''$ SONY GDM-F500R CRT monitor ($1024 \times 768\, px$, $100 Hz$) with a viewable image size of $19.8''$. The display was viewed binocularly and freely from an approximate distance of $132\, cm$, subtending around $17.3 \times 13.0\, deg$ of visual angle for the observer. The monitor was connected to a Wildcat Realizm R500 PCI Express graphics card through a digital video processor (Cambridge Research Systems ViSaGe MKII Stimulus Generator) capable of displaying 14-bit color depths. The monitor was calibrated via a customary software for the stimulus generator (Cambridge Research Systems, Ltd., Rochester, UK) and a ColorCal (Minolta sensor) suction-cup colorimeter.

\subsection{Stimuli}
\label{sec:stimuli}
The software was implemented in Matlab (The MathWorks, Inc.), and the video processor was managed using the Cambridge Research System custom-made toolbox. We used the same spatial configuration of visual stimuli as Otazu et al.~\cite{Otazu2010}, which was inspired by Monnier and Shevell~\cite{Monnier2003,Monnier2004}. In this work, we added a temporal component, following Kaneko and Murakami~\cite{Kaneko2012} flash duration values.

All stimuli were defined in the MacLeod and Boynton color space~\cite{Boynton1986}, which is based in the Smith and Pokorny cone fundamentals~\cite{Smith1975}. In this opponent space, the $l$ axis represents the red-green opponency (\textit{i.e.}, 'L vs M' cone opponency) and the $s$ axis represents the purple-lime opponency (\textit{i.e.}, 'S vs (L+M)' cone opponency), where $s$ is normalized to unity equal-energy white~\cite{Boynton1986}.

\subsubsection{Spatial Configuration}
\label{sec:stimuliSpatialConfiguration}
\begin{figure}[ht]
	\centering
	\includegraphics[width=0.99\linewidth]{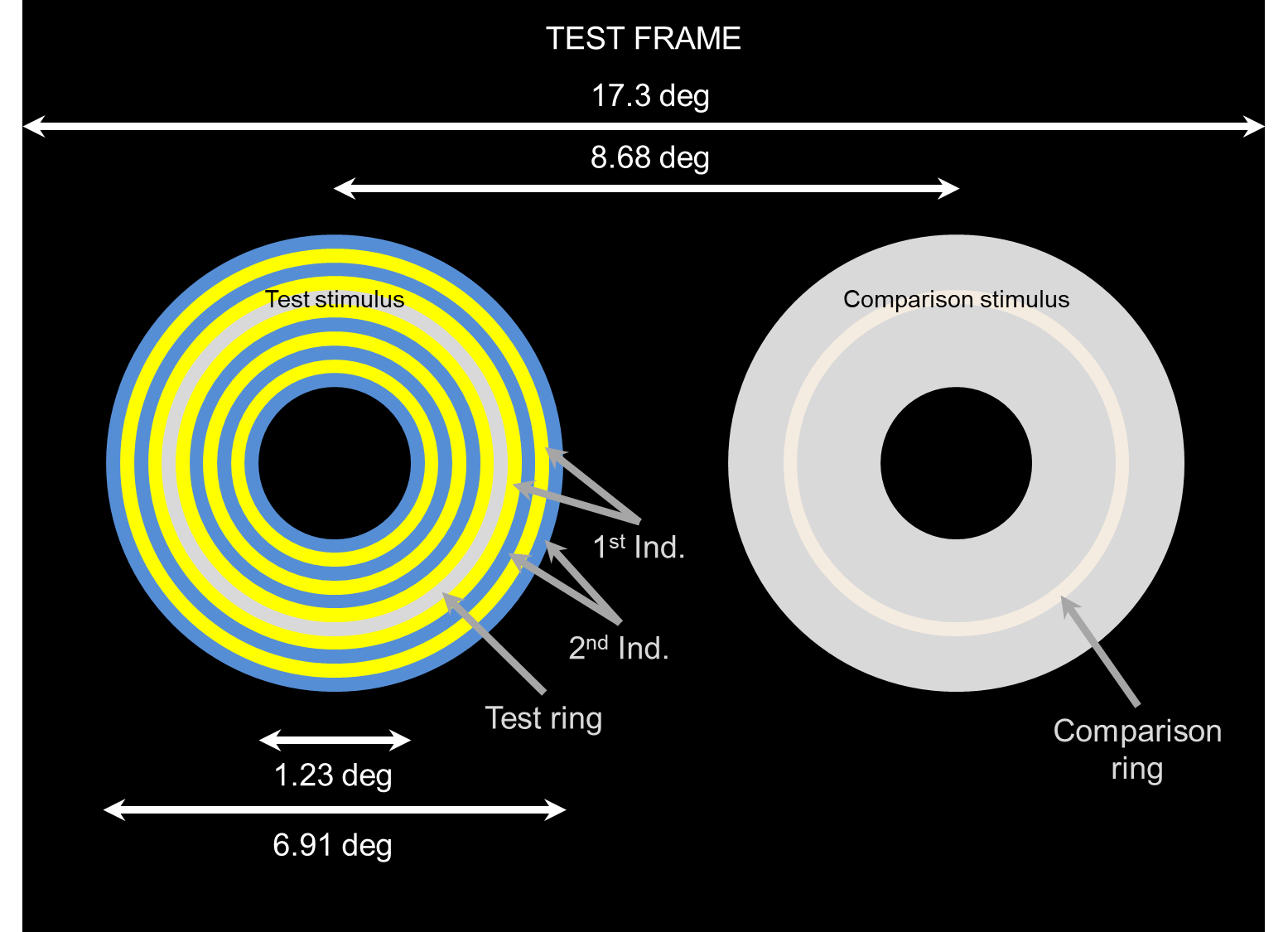}
	\caption{Spatial configuration of the stimuli. Subjects had to adjust the chromaticity of the comparison ring in order to match perceptually the chromaticity of the test ring. The colors in this figure are illustrative. All experimental conditions are described in Table~\ref{table:conditions}.}
	\label{fig:spatialConfig}
\end{figure}

Several stimuli examples are shown in Figure~\ref{fig:spatialConfigExamples} and a schematic of the stimuli's spatial configuration is shown in Figure~\ref{fig:spatialConfig}. The test frame was composed by two circularly symmetric patterns (\textit{i.e.}, the test and the comparison stimuli) presented side by side and separated by $8.68\, deg$ of visual angle from the observer's point of view. Similarly to Otazu et al.~\cite{Otazu2010}, the stimuli's background was dark and the test ring (in the test stimulus) was achromatic and surrounded by concentric rings (inducers) of spatially alternating colors (the 1st and 2nd inducer, according to the physical distance to the test ring). When these two inducers had the same chromaticity, the surround was a uniform region. The striped surround was built with 11 circular stripes (stripes visual frequency was $1.94\, cpd$) because, as observed in Otazu et al.~\cite{Otazu2010}, they produce more color induction than 5 stripes and are not as thin as 17 stripes. In fact, in the 17 stripes case, observers reported that for the shortest flashes they could not detect the test ring. To make the detection of the test ring easier, we drew 8 black dots of 1 pixel size: 4 dots in the inner radius of the ring and 4 points in the outer radius (at 0, 90, 180 and 270\textdegree). The comparison ring, on the right side of the frame, was always surrounded by a uniform achromatic disk approximately metameric to equal-energy white ($l = 0.66$, $s = 0.98$ and $Y = 20\, cd/m^{2}$)~\cite{Monnier2004}.

All chromaticities lay down on the subject's equiluminant plane. The calibration of subjects' equiluminant plane was performed using the Minimally Distinct Border (MDB)~\cite{Boynton1968,Kaiser1971,Wagner1972,Kaiser1990,DeValois1988Chapter,Brill2014,Boynton1973}. Thus, for each subject, all the colors of the stimuli were shown on his/her perceptually equiluminant surface.

\subsubsection{Temporal Configuration}
\label{sec:stimuliTemporalConfiguration}
\begin{figure}[ht]
	\centering
	\includegraphics[width=0.99\linewidth]{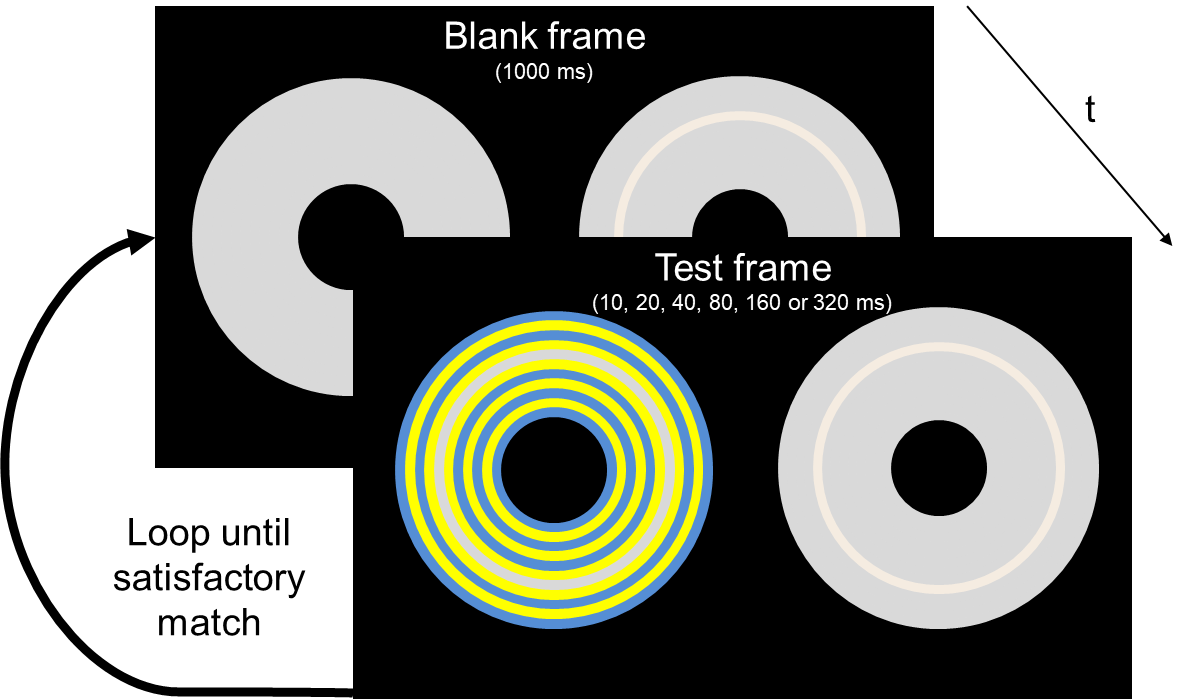}
	\caption{Temporal configuration of the stimuli. The blank frame was shown during $1\,s$ and the duration of the flash (when the test frame was presented) depended on the temporal condition. All experimental conditions are described in Table~\ref{table:conditions}.}
	\label{fig:temporalConfig}
\end{figure}

In Figure~\ref{fig:temporalConfig}, we show the temporal configuration of the stimuli. Two different frames were defined: the test frame and the blank frame. During the blank frame, the test stimulus was an achromatic disk of the same intensity as the test ring. During the test frame, the test stimulus (either striped or uniform) was flashed while the achromatic test ring was not modified. That is, the achromatic test ring remained constant along the experiment and only the chromatic surrounding rings were flashed. The time duration of our blank frame was $1\, s$ and the flash duration took values from $10$ to $320\, ms$ in a dyadic temporal frequency sequence (\textit{i.e.}, $10$, $20$, $40$, $80$, $160$ and $320\, ms$, $N_{tc} = 7$, see temporal conditions in Table~\ref{table:conditions}). The stimulus sequence (\textit{i.e.}, blank and test frames) was repeated until subjects finished the task (see Section~\ref{sec:experimentalProcedure}).

In addition, we also used static stimuli, which was equivalent to infinite flash duration of the test frame. \\

All the experimental conditions (both spatial and temporal) are shown in Table~\ref{table:conditions}. The color of the inducers were chosen to have a good representation of the color space.

\begin{table}[ht!]
    \centering
    \caption{Summary of the experimental conditions (both spatial and temporal). In the spatial conditions, we detail the chromaticity sets in MacLeod and Boynton color space. Only $l$ and $s$ chromatic axes are reported because all chromaticities lay down on the subject's equiluminant plane ($Y = 20\, cd/m^{2}$). In the temporal conditions, we indicate the flash duration in $ms$. During the flash, the test frame was presented. In particular, in the static condition only the test frame was presented (unlimited duration).}
    \label{table:conditions}
    \begin{tabular}{ccccccccc}
        \multicolumn{9}{c}{\textbf{Experiment U} (Uniform surround)} \\
        \multicolumn{9}{c}{Spatial Conditions} \\
        & \multicolumn{2}{c}{Test Ring} & & \multicolumn{2}{c}{1st Inducer} & & \multicolumn{2}{c}{2nd Inducer} \\
        \cline{2-3}\cline{5-6}\cline{8-9}
        & \centering{\textit{l}} & \centering{\textit{s}} & & \textit{l} & \textit{s} & & \textit{l} & \textit{s} \\
        1 & $0.66$ & $0.98$ &  & $0.69$ & $0.98$ &  & $0.69$ & $0.98$ \\
        2 & $0.66$ & $0.98$ &  & $0.66$ & $1.38$ &  & $0.66$ & $1.38$ \\
        3 & $0.66$ & $0.98$ &  & $0.63$ & $0.98$ &  & $0.63$ & $0.98$ \\
        4 & $0.66$ & $0.98$ &  & $0.66$ & $0.58$ &  & $0.66$ & $0.58$ \\
        \multicolumn{9}{c}{Temporal Conditions (in $ms$)} \\
        & 10 & 20 & 40 & 80 & 160 & 320 & static & \\
        \hdashline \\
        \multicolumn{9}{c}{\textbf{Experiment S} (Striped surround)} \\
        \multicolumn{9}{c}{Spatial Conditions} \\
        & \multicolumn{2}{c}{Test Ring} & & \multicolumn{2}{c}{1st Inducer} & & \multicolumn{2}{c}{2nd Inducer} \\
        \cline{2-3}\cline{5-6}\cline{8-9}
        & \centering{\textit{l}} & \centering{\textit{s}} & & \textit{l} & \textit{s} & & \textit{l} & \textit{s} \\
        1 & $0.66$ & $0.98$ &  & $0.69$ & $0.98$ &  & $0.63$ & $0.98$ \\
        2 & $0.66$ & $0.98$ &  & $0.66$ & $1.38$ &  & $0.66$ & $0.58$ \\
        3 & $0.66$ & $0.98$ &  & $0.63$ & $0.98$ &  & $0.69$ & $0.98$ \\
        4 & $0.66$ & $0.98$ &  & $0.66$ & $0.58$ &  & $0.66$ & $1.38$ \\
        \multicolumn{9}{c}{Temporal Conditions (in $ms$)} \\
        & 10 & 20 & 40 & 80 & 160 & 320 & static & \\
    \end{tabular}
\end{table}

\subsection{Subjects}

The experiments were done by six observers ($N_{sub} = 6$), four of them from our lab (AA, DB, LR and XO), who were familiar with color spaces, and two others who were not related to the lab (BG and CM). All of them were informed of the aim of the experiments and consented to participate in the experimentation. Five observers were completely na{\"\i}ve (AA, BG, CM, DB and LR), while one of them is one of the authors of the paper (XO). Three of them are male (AA, DB and XO) and the other three are female (BG, CM and LR). The ages were comprised between 22 and 45 years old. All of them had normal or corrected-to-normal vision, tested using the Ishihara~\cite{Ishihara2003} and the D-15 Farnsworth Dichotomous Test~\cite{Farnsworth1947}. To learn the experimental procedure, all observers did a one day training session before starting the experiments.

\subsection{Experimental Procedure}
\label{sec:experimentalProcedure}

The subjects' task was to adjust the chromaticity of the comparison ring until they perceived it equal to the test ring. They were instructed to do the matching according to the test ring color perceived during the test frame, ignoring the after-effect produced during the blank frame.

We conducted two different experiments (Experiments U and S) in accordance with the Code of Ethics of the World Medical Association (Declaration of Helsinki) to study how the surround chromaticity and the flash duration influence the color perception. Since color induction strongly depends on the surround type, we divided the experiments according to it. In Experiment U, the test ring had a uniform surround and, in Experiment S, the test ring had a striped surround. 

To reduce the available color space to one-dimension, similarly to Kaneko and Murakami~\cite{Kaneko2012}, we performed a previous experiment where subjects were able to adjust the chromaticity in a MacLeod and Boynton two-dimensional color space. We observed that in both experiments (Experiment U and S) the observations were approximately on the cardinal axis which includes the test and inducers' chromaticities. Thus, we reduced the available color space to one-dimension, \textit{i.e.}, the observers only changed the comparison ring chromaticity along $l$ or $s$ axes, depending on the experimental spatial condition (1-D Experiment).

For each 1-D Experiment (Experiments U and S) we had 28 different experimental conditions (4 different spatial conditions -$N_{sc} = 4$- and 7 temporal conditions -$N_{tc} = 7$-, see Table~\ref{table:conditions}). Each subject evaluated each experimental condition 10 different times (\textit{i.e.}, 10 different observations, $N_{obs} = 10$). An experimental condition was determined by the combination of a spatial condition and a temporal condition. Each run started with 3 minutes of dark adaptation~\cite{Kaneko2012} and subjects performed 48 different observations (8 different spatial conditions, 3 different temporal conditions, and 2 repetitions). Subjects did not have any time restriction, but they were advised not to take more than 1 minute for each experimental condition. On average, each run took about 30 minutes. Subjects evaluated the static conditions of both experiments on two days apart. In that case, they performed 24 observations in each run (8 different spatial conditions, 1 temporal condition -static- and 3 repetitions, except for the last run, which was 4 repetitions), taking about 15 minutes.

To minimize memory effect of the experiments, we defined a pseudo-random order of the temporal conditions in each run, while the spatial conditions were randomized.

\section{Results}

\subsection{Metric}
\label{sec:metric}
To represent the induction strength, we define the perceptual color induction metric

\begin{equation}
    \Delta C_{i} = \frac{C^{c}_{i}-C^{t}_{i}}{C^{s}_{i}-C^{t}_{i}}\;,
    \label{eq:metric}
\end{equation}

\noindent
where $i$ is the chromatic axis of the MacLeod-Boynton color space along which the induction was measured ($i = [l, s]$). $C^{c}_{i}$ is the scalar value of the $i$th axis in the lsY color space of the comparison ring (the observation done by the subject); $C^{t}_{i}$ is the scalar value of the $i$th axis of the test ring; and $C^{s}_{i}$ is the scalar value of the $i$th axis of the 1st inducer. Note that, since it is a 1-D Experiment, all the terms of the equation are scalar because it only considers the axis where the chromaticity could be adjusted. This metric is sensitive to both color contrast and color assimilation: when $\Delta C$ is positive, color assimilation is induced because $C^{c}_{i}$ is shifted towards $C^{s}_{i}$; and when $\Delta C$ is negative, color contrast is induced because $C^{t}_{i}$ is shifted away from $C^{s}_{i}$. Nevertheless, $|\Delta C|$ has to be greater than the just noticeable difference (JND) to consider that color induction is produced. The JND region was computed in the CIELab color space ($\Delta E = 1$), which is an approximately perceptually uniform color space, and then transformed to the MacLeod-Boynton color space. Thus, when $-JND < \Delta C < JND$ we consider that there is no induction. This metric does not include the 2nd inducer because only the 1st inducer determines the type of color induction (color contrast or color assimilation).

For each experimental condition, we averaged all the 10 observations of each subject and computed the average and the standard error of means (SEM) of all 6 subjects ($N_{sub} = 6$).

\subsection{Experiment U}
\begin{figure*}[htb]
	\centering
	\includegraphics[height=0.49\textheight]{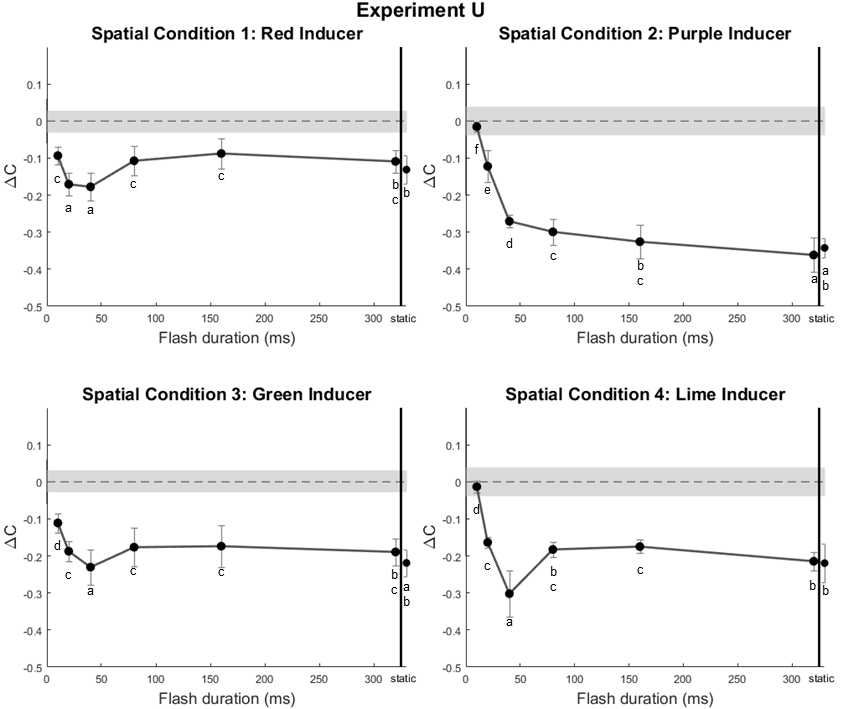}
	\caption{Results of Experiment U. The averaged (mean of 6 observers) color induction metric ($\Delta C$) is plotted against the flash duration (7 different temporal conditions). Separate plots correspond to the different spatial conditions (see Experiment U in Table~\ref{table:conditions}). The gray area is the region where subjects could not perceive chromatic differences ($\Delta C \in [-JND,JND]$). Error bars indicate $\pm 1\, SEM$. An ANOVA analysis of the data (see Table~\ref{table:statistics}) showed that there were significant differences in color induction strength for different temporal conditions. Fisher's post-hoc analysis (letters below the error bars), which allowed us to measure which temporal conditions were significantly different, stressed that the peak of color induction was always perceived at $40\, ms$, except for Spatial Condition 2. Furthermore, static stimulation induced the same color as the longest flash duration ($320\, ms$) and the perceived color at $80$ and $160\, ms$ did not vary. In all chromatic conditions, chromatic contrast ($\Delta C<0$) or no induction was induced.}
	\label{fig:Results1DUniform}
\end{figure*}

In this experiment, we studied the color induced by equiluminant uniform surrounds on an achromatic test ring at different temporal conditions (see Table~\ref{table:conditions}). The results (Figure~\ref{fig:Results1DUniform}) show that chromatic contrast was induced in all experimental conditions, except at the $10\, ms$ flash in the purple-lime opponent axis. In fact, under these experimental conditions (Spatial Conditions 2 and 4 -purple and lime inducers- flashed during $10\, ms$), observers pointed out that the test frame detection was very difficult.

We used a nested ANOVA analysis, which had 6 and 378 degrees of freedom ($N_{tc}-1$ and $N_{tc}N_{sub}(N_{obs}-1)$, respectively) to observe whether temporal conditions affected to color induction. Once the nested ANOVA indicated that there were significant differences, a Fisher's Least Significant Differences post-hoc analysis (Fisher's LSD) was performed to group the temporal conditions according to the color induction they induced. These analyses showed that for Spatial Conditions 1, 3 and 4, maximum color contrast induction was produced by $40\, ms$ flash, while in the Spatial Condition 2 there is no peak at $40\, ms$. Furthermore, for all spatial conditions there are no significant differences between the perceived colors at $80$ and $160\, ms$. Moreover, the induction produced by a $320\, ms$ flash was similar to the induction produced by a static stimulus. All the ANOVA statistics' details are shown in Table~\ref{table:statistics} and the letters below error bars in Figure~\ref{fig:Results1DUniform} show the temporal conditions that induced statistically similar colors (Fisher's LSD post-hoc analysis' results). The temporal conditions which have the same letter can be considered that induce the same perceptual color.

\subsection{Experiment S}
\begin{figure*}[t]
	\centering
	\includegraphics[height=0.49\textheight]{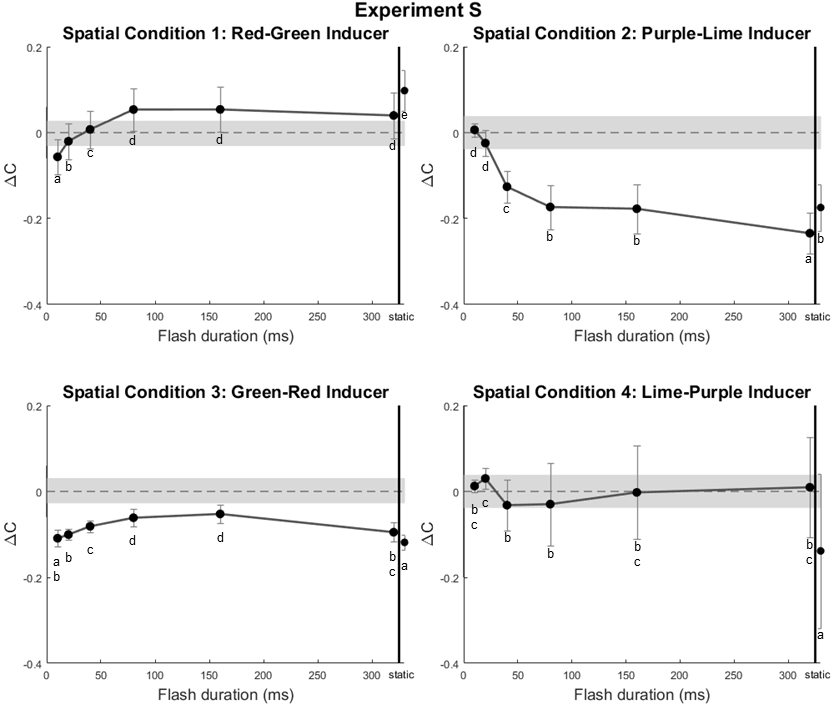}
	\caption{Results of Experiment S. The averaged (mean of 6 observers) color induction metric ($\Delta C$) is plotted against the flash duration. Separate plots correspond to the different spatial conditions (see Experiment S in Table~\ref{table:conditions}). The gray region indicates where induction could not be perceived ($[-JND, JND]$). Error bars indicate $\pm 1\, SEM$. An ANOVA analysis of the data (see Experiment S in Table~\ref{table:statistics}) showed that there were significant differences between color induction for different temporal conditions. Like in Experiment U, we did a Fisher's LSD post-hoc analysis (letters below error bars) to know the temporal conditions that induced significantly different colors. In this Experiment, we observed that the static stimuli induction had significant differences with respect to the induction produced by a $320\, ms$ flash. No color assimilation had been induced, except by static stimulus in Spatial Condition 1.}
	\label{fig:Results1DStriped}
\end{figure*}

In this experiment, we studied the color induced by equiluminant striped surrounds on an achromatic test ring under different temporal conditions (see Experiment S in Table~\ref{table:conditions}). In Figure~\ref{fig:Results1DStriped}, we can observe that Spatial Conditions 1 and 4 did not induce any color induction when the stimuli were flashed. Only one out of 6 subjects observed assimilation in flashed Spatial Condition 1, and 2 out of 6 perceived assimilation in flashed Spatial Condition 4. By contrast, static stimulus of Spatial Condition 1 induced chromatic assimilation ($\Delta C>0$). In Spatial Conditions 2 and 3 only color contrast ($\Delta C<0$) was perceived.

Subjects pointed out that, again, they left a gray color for Spatial Conditions 2 and 4 (purple-lime axis) flashed during 10 ms because they were not able to see the test frame.

Similarly to Experiment U, ANOVA analysis showed significant differences between the induction produced by flashes of different durations in all spatial conditions (see Experiment S in Table~\ref{table:statistics}). We have not observed any grouping or behavioral pattern over all spatial conditions in the Fisher's LSD post-hoc analysis. Since the profile of the results of Figures~\ref{fig:Results1DUniform} and~\ref{fig:Results1DStriped} are completely different and the strongest induction in this experiment is not observed at the same flash duration as in the previous one, our initial hypothesis should be rejected.

\begin{table}[ht!]
    \centering
    \caption{Summary of the nested ANOVA results. These results show that, in all spatial conditions, color induction depended on the temporal condition of the stimulus (\textit{i.e.}, the flash duration). The nested ANOVAs had 6 and 378 degrees of freedom, corresponding to $N_{tc}-1$ and $N_{tc}N_{sub}(N_{obs}-1)$, respectively.}
    \label{table:statistics}
    \begin{tabular}{ccc}
        \multicolumn{3}{c}{\textbf{Experiment U} (Uniform surround)} \\
        Spatial & \multirow{2}{*}{$F_{6,378}$} & \multirow{2}{*}{p} \\
        Condition & & \\
        1 & $18.74$     & $<0.001$ \\
        2 & $160.80$    & $<0.001$ \\
        3 & $11.95$     & $<0.001$ \\
        4 & $41.43$     & $<0.001$ \\
        \hdashline \\
        \multicolumn{3}{c}{\textbf{Experiment S} (Striped surround)} \\
        Spatial & \multirow{2}{*}{$F_{6,378}$} & \multirow{2}{*}{p} \\
        Condition & & \\
        1 & $84.26$     & $<0.001$ \\
        2 & $27.19$     & $<0.001$ \\
        3 & $20.56$     & $<0.001$ \\
        4 & $10.88$     & $<0.001$ \\
    \end{tabular}
\end{table}

\section{Discussion}
We have divided the discussion into two parts, according to the two experiments (Experiment U and Experiment S).

\subsection{Uniform surround (Experiment U)}
Our results from uniform surround stimuli (see Figure~\ref{fig:Results1DUniform}) show that static uniform surrounds induced color contrast in all experimental conditions, in line with previous works~\cite{Monnier2003,Monnier2004,Gordon2006,Otazu2010,Kaneko2012}. Concretely, Gordon and Shapley~\cite{Gordon2006} used uniform surrounds to study how the luminance and brightness of the test region affect color induction, observing color contrast in all conditions. In accordance to Kirshcmann's Third Law~\cite{Kirschmann1891}, they concluded that brightness, but not luminance, is crucial in the effect on color induction. They found that color contrast is maximal when the stimulus is equibrightness (but not equiluminant) and, as brightness contrast is increased, the color contrast is reduced. In their work, Kaneko and Murakami~\cite{Kaneko2012} also used equiluminant stimuli (using the heterochromatic flicker photometry technique) and they also found color contrast under all conditions. They used flashed stimuli and measured the color contrast induced by color surrounds presented in different temporal conditions. They concluded that the shorter the flash duration, the stronger the color contrast. By contrast, we observed a clear peak of color contrast when stimuli were flashed during $40\, ms$. Despite our work shares several features with Kaneko and Murakami~\cite{Kaneko2012} such as temporal conditions, temporal configuration, equiluminant stimuli and methodology, there are some differences between the current study and their work. They measured color induction on a central disk and we measured it on a ring similar to the one used by Monnier and Shevell~\cite{Monnier2003} and Otazu et al.~\cite{Otazu2010}. Thus, the visual angle of the evaluated feature was different (a disk of $1\, deg$ and a ring width of $15.5\, min$ of visual angle). In addition, they introduced a thin black ring around the central disk, \textit{i.e.}, a border of lower luminance between the central disk and its color surround, which could lead to different results~\cite{Xing2015}. Carefully analyzing their raw data, we can observe that in some spatial conditions there is not a clear peak of color contrast at the shortest flash ($10\, ms$), but there is around $20$ and $40\, ms$. Moreover, they showed two subjects (subjects MS and YY) who seemed to obtain similar results to ours: they had a peak of induction during the short flashes (around $20\, ms$), but not at the shortest ($10\, ms$). Thus, all these reasons could explain the dissimilarity between our and their results.

It is assumed that color induction (both color contrast and assimilation) is the result of neural mechanisms in V1~\cite{Zaidi1992,Zaidi1999,DeWeert1997,Cao2005} and stimuli on $l$ (red and green surrounds) and $s$ (purple and lime surrounds) axes are nearly independently processed at the first stages of the Human Visual System (HVS), \textit{i.e.}, in the retina, LGN and V1~\cite{Sincich2005}. Since stimuli on the $l$ axis are processed by the parvocellular pathway and stimuli on $s$ axis are processed by the koniocellular pathway~\cite{Nassi2009}, we expected to obtain different results using stimuli that independently activate these different visual pathways. In particular, we expected to observe different temporal behaviors for each pathway because parvocellular and koniocellular pathways have different processing speeds~\cite{Casagrande2007,Briggs2009}. From the obtained results (see Figure~\ref{fig:Results1DUniform}), we can see that the color contrast when the inducer is purple (Spatial Condition 2 of Table~\ref{table:conditions}) is completely different to that with other color surrounds. This spatial condition induced color contrast, \textit{i.e.}, the achromatic test ring was perceived as lime, when flashed longer than $20\, ms$. Moreover, it is the only color of the surround with no induction maximum, except at infinite flash duration (\textit{i.e.}, static temporal condition). This stimulus activates the S-OFF channel of koniocellular pathway, which directly projects to layer 4A of V1~\cite{Chatterjee2003,Callaway2014}. By contrast, the S-ON channel of the koniocellular, and both parvocellular and magnocellular pathways project to layers 2/3, 4C$\beta$, and 4C$\alpha$ of V1, respectively, and all of them converge into layer 2/3 of V1~\cite{Sincich2005}. The different processing of S-ON and S-OFF channels of koniocellular pathway could explain the dissimilar psychophysical results (see Figure~\ref{fig:Results1DUniform}) on color contrast when inducers had lime (S-ON on test ring) or purple (S-OFF on test ring) chromaticities. In addition, since all channels of parvocellular pathway (L-ON, L-OFF, M-ON, and M-OFF) are processed in the same layers of V1 (first in layer 4C$\beta$, 4B and finally in 2/3), and S-ON channel of koniocellular pathway is mainly processed in layer 2/3, it could explain this similarity in color induction when inducers had red, green or lime chromaticities (Spatial Conditions 1, 2 and 4).

\subsection{Striped surround (Experiment S)}
Our results from striped surround stimuli (see Figure~\ref{fig:Results1DStriped}) show that, similarly to other authors~\cite{Fach1986,Smith2001}, striped surrounds can induce color contrast, but we have not observed it under all experimental conditions. Only one out of 28 experimental conditions induced color assimilation, namely when the stimulus was static, its 1st inducer was red and the 2nd one was green. In contrast, Monnier and Shevell~\cite{Monnier2003,Monnier2004} and Otazu et al.~\cite{Otazu2010} observed that striped surrounds induce color assimilation in all spatial conditions and never found color contrast. Although our work has very similar features to these works, such as spatial configuration and chromaticities (see Section~\ref{sec:stimuli}), they used non-equiluminant stimuli and we used equiluminant stimuli. Thus, our hypothesis is that this luminance difference explains the difference between our results and the ones obtained by these authors. Monnier and Shevell~\cite{Monnier2003,Monnier2004}, and Otazu et al.~\cite{Otazu2010} introduced a luminance difference between the test ring and its surround, and they found chromatic assimilation for striped surrounds. De Weert and Spillmann~\cite{DeWeert1995} found no color induction in equiluminant stimuli, but found color induction when the 1st inducer had lower luminance than the central region. Extending this work, Cao and Shevell~\cite{Cao2005} showed that assimilation in the $l$ axis of MacLeod and Boynton color space~\cite{Boynton1986} was found when the inducer luminance was lower than the central region luminance, but not when it was higher, observation that was also reported by De Weert and Spillmann~\cite{DeWeert1995}. In the $s$ axis, they showed that color assimilation does not depend on the inducing luminance (\textit{i.e.}, induction was observed when the inducing luminance was either lower or higher than the central region luminance), but depends on the spatial configuration of the inducers (\textit{i.e.}, on both spatial frequency and inducer's spatial separation). Thus, our hypothesis is supported by De Weert and Spillmann~\cite{DeWeert1995} and Cao and Shevell~\cite{Cao2005} results. Considering that chromatic assimilation mainly appears when stimulus is not equiluminant, \textit{i.e.}, only when magnocellular pathway is activated, it could suggest that magnocellular pathway could act as a switch-like signaling system activating or deactivating assimilation in both parvocellular and koniocellular pathways in layer 2/3 (where all the pathways converge). In recent neurophysiological and psychophysical studies~\cite{Xing2015,nunez2018}, the authors concluded that brightness and color interact in V1. In particular, their work supported the hypothesis that the color appearance depends on brightness contrast~\cite{Kirschmann1891,Gordon2006,Faul2008,Bimler2009} because there is a mutual-suppression (\textit{i.e.}, color assimilation) between color-responsive cells and luminance-responsive cells. Furthermore, they proposed that these interactions are driven by double-opponent cells, which respond to both luminance and color differences (Color-Lum neurons)~\cite{Johnson2001,Johnson2008}. The non-opponent neurons, or Lum neurons, are inactive when the stimulus is equiluminant, and single-opponent neurons, or Color neurons, respond to large areas of color and do not respond to luminance differences. These neurophysiological observations, also supports our hypothesis that luminance difference between the test ring and its surround, which activates Lum neurons, could be a key factor to induce color assimilation. In addition, psychophysical studies by Fach and Sharpe~\cite{Fach1986}, and Smith, Jin and Pokorny~\cite{Smith2001} concluded that spatial frequency is another crucial factor to induce color assimilation in equiluminant striped stimuli. In particular, they observed that very thin stripes ($>9\, cpd$) induce color assimilation and thick stripes ($<0.7\, cpd$) induce color contrast, with a transition point from assimilation to contrast around $4\, cpd$~\cite{Smith2001}. Considering that the spatial frequency of our stimuli is $1.94\, cpd$, we agree with them~\cite{Fach1986,Smith2001,Cao2005}: in that range both color contrast and color assimilation could be induced.

Comparing the results from flashed and static stimuli, we can observe that in almost all spatial conditions (except for Spatial Condition 1), the type of color induction, \textit{e.g.} assimilation or contrast, did not vary between flashed and static stimuli, when the flash was longer than $40\, ms$.\\

In both types of surround configurations, \textit{e.g.}, uniform and striped surrounds (see Figures~\ref{fig:Results1DUniform} and~\ref{fig:Results1DStriped}), we can see that at the shortest flash duration ($10\, ms$) subjects only perceived color induction when the surrounding colors were on the $l$ axis of MacLeod and Boynton color space, but did not perceive any induction when the surround was on the $s$ axis (purple or lime colors). In fact, when subjects finished the experiment they pointed out that under these experimental conditions (purple and lime surround colors flashed for $10\, ms$) they were not able to see the test frame and, therefore, left an achromatic comparison ring. This consideration goes in line with the physiological observation that the koniocellular pathway is slower than the parvocellular pathway~\cite{Casagrande2007,Briggs2009}, \textit{i.e.}, the purple and lime colors are processed more slowly than the red and green colors.

\section{Conclusions}
Taking into account that we only observed color contrast (except for red-green inducer in static striped stimuli) for the two different types of surround (uniform and striped) and that temporal behavior of color induction depends on visual pathways (see Figures~\ref{fig:Results1DUniform} and~\ref{fig:Results1DStriped}), we can conclude:

\begin{itemize}
    \item The strongest color contrast is induced by a uniform surround stimulus flashed for $40\, ms$.
    \item Purple inducer (which induces a lime chromaticity and, thus, activates the S-OFF channel in layer 4A) induces a temporal response that is completely different to the temporal response induced by other inducers such as red, green and lime (which activates other channels in layer 2/3).
    \item Striped equiluminant stimuli do not induce color assimilation (except for red-green inducers).
    \item The test frame cannot be perceived during flashes shorter than $20\, ms$ when the colors of the surrounding are on the \textit{s} axis of the MacLeod and Boynton color space.
    \item Our initial hypothesis, \textit{i.e.}, flashed uniform and striped surrounds would induce opposite colors but with a similar temporal behavior, should be rejected.
\end{itemize}
    
Considering previous studies, we can also conclude that luminance could be a key factor for color assimilation. In particular, assimilation only appears in non-equiluminant stimuli, or in equiluminant striped stimuli with a very high spatial frequency. This could suggest that color contrast and color assimilation effects are the result of different mechanisms or, at least, the result of the same mechanism which needs an interaction with luminance to induce color assimilation.

\section{Future Work}
A more detailed study of the specific contribution of luminance to color assimilation is needed. There exist some works in this direction~\cite{DeWeert1995,Cao2005}, but they did not perform a systematic study of color induction depending on luminance differences (they used few arbitrary luminance values, and Cao and Shevell~\cite{Cao2005} did not use equiluminant stimuli). In particular, it is interesting to study how the $\Delta C$ values (see Section~\ref{sec:metric}) change depending on test ring, 1st and 2nd inducers' luminance differences. It would help to study the precise dependence of chromatic induction on both equiluminant and non-equiluminant stimuli. It would help to unify the current work and several previous works~\cite{Monnier2003,Monnier2004,Otazu2010} under a common framework.

\section{Funding Information}

This work is partially supported by: Spanish Ministry of Economy, Industry and Competitiveness (DPI2017-89867-C2-1-R); Agencia de Gestio d'Ajuts Universitaris i de Recerca (AGAUR) (2017-SGR-649); CERCA Programme / Generalitat de Catalunya.

\section{Acknowledgements}
The authors would like to thank Javier Retana for his useful comments on statistical analysis procedures, Olivier Penacchio for his comments on the work and C. Alejandro Parraga for his comments on the laboratory setup. The authors also want to thank all the subjects for their valuable time.

\bibliographystyle{plain}
\bibliography{main}
\end{document}